\documentclass[a4paper]{jpconf}

\usepackage[T1]{fontenc}

\usepackage{graphicx}

\usepackage{amsmath, amssymb, amsfonts}
\usepackage{braket}
\usepackage{color}

\usepackage{textpos}
\usepackage{etoolbox}
\patchcmd{\thebibliography}{\advance\leftmargin\labelsep}
  {\labelsep=0.5cm \advance\leftmargin\labelsep}{}{}
  
\begin{document}

\begin{textblock}{4}(12,-2)
\begin{flushright}
\begin{footnotesize}
22 March 2019
\end{footnotesize}
\end{flushright}
\end{textblock}

\title{Symmetry properties of non-Hermitian $\mathcal{PT}$-symmetric quantum field theories}

\author{P Millington}

\address{School of Physics and Astronomy, University of Nottingham, \\ Nottingham NG7 2RD, UK}

\ead{p.millington@nottingham.ac.uk}

\begin{abstract}
We describe recent progress in understanding the continuous symmetry properties of non-Hermitian, $\mathcal{PT}$-symmetric quantum field theories. Focussing on a simple non-Hermitian theory composed of one complex scalar and one complex pseudoscalar, we revisit the derivation of Noether's theorem to show that the conserved currents of non-Hermitian theories correspond to transformations that do not leave the Lagrangian invariant. We illustrate the impact that this has on the consistent formulation of (Abelian) gauge theories by studying a non-Hermitian extension of scalar quantum electrodynamics. We consider the spontaneous breakdown of both global and local symmetries, and describe how the Goldstone theorem and the Englert-Brout-Higgs mechanism are borne out for non-Hermitian, $\mathcal{PT}$-symmetric theories.
\end{abstract}

\section{Introduction}

The standard lore of quantum mechanics is that operators corresponding to real-valued observables must be Hermitian. However, not all matrices with real eigenvalues are Hermitian, and, in the case of the Hamiltonian, it turns out that the reality of the eigenspectrum~\cite{Bender:1998ke}, and unitary evolution~\cite{Bender:2002vv}, can instead by guaranteed by the weaker condition of unbroken $\mathcal{PT}$ symmetry, that is symmetry under the combined action of parity $\mathcal{P}$ and time-reversal $\mathcal{T}$ transformations (for overviews of $\mathcal{PT}$-symmetric quantum mechanics~\cite{Bender:1998gh}, see references~\cite{Bender:2005tb, Bender:2007nj}).

In this talk, we consider the continuous symmetry properties of non-Hermitian, $\mathcal{PT}$ symmetric field theories, summarising the results of references~\cite{Alexandre:2017foi, Alexandre:2018uol, Alexandre:2018xyy} (see also reference~\cite{Mannheim:2018dur}) and focussing, in particular, on how Noether's theorem~\cite{Noether}, the Goldstone theorem~\cite{Nambu:1960xd, Goldstone:1961eq, Goldstone:1962es} and the Englert-Brout-Higgs mechanism~\cite{Englert:1964et, Higgs:1964ia, Higgs:1964pj} are borne out. In the context of a complex scalar model, we show that there exist conserved currents for non-Hermitian theories, but the corresponding transformations do not leave the Lagrangian invariant~\cite{Alexandre:2017foi}. In the case of spontaneously broken global symmetries, the existence of the conserved current is sufficient to ensure that Goldstone's theorem still holds~\cite{Alexandre:2018uol}, and we obtain a massless Goldstone mode. However, in the case of gauge theories, coupling minimally to the conserved current means the Lagrangian is not gauge invariant~\cite{Alexandre:2018xyy}. As a result, we must couple to a non-conserved current, and, in the case of non-Hermitian scalar quantum electrodynamics, the consistency of the Maxwell equations precludes, in general, our working in Lorenz gauge (i.e.~setting the four-divergence of the gauge field to zero)~\cite{Alexandre:2018xyy}. With these subtleties understood, we find that the Englert-Brout-Higgs mechanism can still generate a gauge-invariant mass for the vector boson~\cite{Alexandre:2018xyy}. Throughout what follows, we focus only on the regimes in which the eigenspectra remain real and avoid the exceptional points of the theories (see, e.g., reference~\cite{Heiss:2012dx}); for a complementary discussion, see reference~\cite{Mannheim:2018dur}.

\section{A scalar model}
\label{sec:model}

We consider the following non-Hermitian complex-scalar theory with $\mathcal{PT}$ symmetry~\cite{Alexandre:2017foi}:
\begin{equation}
\label{eq:Lag1}
\mathcal{L}\ =\ \partial_{\alpha}\phi_1^*\partial^{\alpha}\phi_1\:+\:\partial_{\alpha}\phi_2^*\partial^{\alpha}\phi_2\:-\:m_1^2|\phi_1|^2\:-\:m_2^2|\phi_2|^2\:-\:\mu^2(\phi_1^*\phi_2-\phi_2^*\phi_1)\;,
\end{equation}
where $m_1^2,m_2^2,\mu^2\geq 0$. It is $\mathcal{PT}$-symmetric if the c-number fields $\phi_1$ and $\phi_2$ transform as
\begin{equation}
\mathcal{P}:\ \begin{pmatrix} \phi_1 \\ \phi_2\end{pmatrix}\ \rightarrow\ \begin{pmatrix} +\,\phi_1 \\ -\,\phi_2\end{pmatrix}\qquad \text{and} \qquad \mathcal{T}:\ \begin{pmatrix} \phi_1 \\ \phi_2\end{pmatrix}\ \rightarrow\ \begin{pmatrix} +\,\phi_1^* \\ +\,\phi_2^*\end{pmatrix}\;.
\end{equation}
Notice that we have chosen $\phi_2$ to transform as a pseudoscalar. For a discussion of the discrete symmetry properties of the corresponding operators in Fock space, see, e.g., reference~\cite{Mannheim:2018dur}.

For $|m_1^2-m_2^2|\geq 2\mu^2$, the squared mass eigenvalues
\begin{equation}
\label{eq:eigenspectrum}
M_{\pm}^2\ =\ \frac{1}{2}(m_1^2+m_2^2)\:\pm\:\frac{1}{2}\sqrt{(m_1^2-m_2^2)^2-4\mu^4}
\end{equation}
are real, and the theory is in the unbroken phase of $\mathcal{PT}$ symmetry. The eigenvectors of the mass matrix are \smash{$\mathbf{e}_+\propto(M_+^2-m_2^2,-\mu^2)^{\mathsf{T}}$} and \smash{$\mathbf{e}_-\propto (M_-^2-m_2^2,-\mu^2)^{\mathsf{T}}$}. They are not orthogonal with respect to Hermitian conjugation, i.e.~\smash{$\mathbf{e}_{\pm}^{\dagger}\mathbf{e}_{\mp}\neq 0$}, but they are orthogonal with respect to $\mathcal{PT}$-conjugation, i.e.~\smash{$\mathbf{e}_{\pm}^{\ddagger}\mathbf{e}_{\mp}=0$}, where \smash{$\mathbf{e}^{\ddagger}\equiv(P\mathbf{e}^{*})^{\mathsf{T}}$} with $P\equiv{\rm diag}(1,-1)$.\footnote{To the best of our knowledge, the $\ddagger$ notation was first introduced in reference~\cite{Bender:1998gh} for the $\mathcal{PT}$ conjugate. The notation was extended in reference~\cite{Alexandre:2017foi} to include matrix transposition, denoted here by a superscript $\mathsf{T}$.}

However, the action is not Hermitian, and it turns out that we cannot simultaneously satisfy
\begin{equation}
\frac{\partial \mathcal{L}}{\partial \Phi^{\dagger}}\:-\:\partial_{\alpha}\,\frac{\partial \mathcal{L}}{\partial (\partial_{\alpha}\Phi^{\dagger})}\ =\ 0\quad \text{and}\quad \frac{\partial \mathcal{L}}{\partial \Phi}\:-\:\partial_{\alpha}\,\frac{\partial \mathcal{L}}{\partial (\partial_{\alpha}\Phi)}\ =\ 0\;,\quad \text{where}\quad \Phi\ \equiv\ \begin{pmatrix} \phi_1 \\ \phi_2\end{pmatrix}\;,
\end{equation}
except for the trivial solution $\phi_1=\phi_2=0$~\cite{Alexandre:2017foi}. This is just the statement that the left and right eigenspectra of non-Hermitian matrices are, in general, distinct. We are nevertheless free to choose in which of these the zero mode resides. In other words, we can choose one of the usual Euler-Lagrange equations to define the equations of motion (see section~\ref{sec:Noether}). The two choices are, however, physically equivalent, since the difference amounts to a sign change on the non-Hermitian terms ($\mu^2\to-\,\mu^2$), which can be absorbed into a field redefinition.

Choosing to define the equations of motion by the variation with respect to \smash{$\Phi^{\ddagger}\equiv (P\Phi^*)^{\mathsf{T}}$} (or, equivalently, with respect to $\Phi^{\dagger}$), we have
\begin{equation}
(\Box+m_1^2)\phi_1+\mu^2\phi_2\ =\ 0\qquad \text{and} \qquad
(\Box+m_2^2)\phi_2-\mu^2\phi_1\ =\ 0\;.
\end{equation}
Notice that these classical equations of motion are \emph{not} $\mathcal{PT}$-symmetric, such that non-trivial solutions will, in general, break the $\mathcal{PT}$ symmetry spontaneously (see section~\ref{sec:Goldstone}).

\section{Noether's theorem}
\label{sec:Noether}

Turning now to the variational procedure, the variation of the action is
\begin{align}
\delta S\ &=\ \int{\rm d}^4x\;\bigg[\bigg(\frac{\partial \mathcal{L}}{\partial \Phi}-\partial_{\alpha}\,\frac{\partial \mathcal{L}}{\partial (\partial_{\alpha}\Phi)}\bigg)\delta \Phi\:+\:\delta \Phi^{\ddagger}\bigg(\frac{\partial \mathcal{L}}{\partial \Phi^{\ddagger}}-\partial_{\alpha}\,\frac{\partial \mathcal{L}}{\partial (\partial_{\alpha}\Phi^{\ddagger})}\bigg)\bigg]\nonumber \\&\qquad+\partial_{\alpha}\bigg(\frac{\partial \mathcal{L}}{\partial (\partial_{\alpha}\Phi)}\,\delta \Phi+\delta \Phi^{\ddagger}\,\frac{\partial \mathcal{L}}{\partial (\partial_{\alpha}\Phi^{\ddagger})}\bigg)\bigg]\;.
\end{align}
Since only one of the Euler-Lagrange equations can, in general, be satisfied, requiring $\delta S=0$ means that the other must be supported by an external source or non-vanishing surface terms~\cite{Alexandre:2017foi}.

Alternatively, we see that the Noether current
\begin{equation}
j^{\alpha}_{\delta}\ =\ \frac{\partial \mathcal{L}}{\partial (\partial_{\alpha}\Phi)}\,\delta \Phi+\delta \Phi^{\ddagger}\,\frac{\partial \mathcal{L}}{\partial (\partial_{\alpha}\Phi^{\ddagger})}
\end{equation}
is conserved only if it corresponds to a transformation that effects a particular variation of the Lagrangian; that resulting from the non-vanishing of the other Euler-Lagrange equation~\cite{Alexandre:2017foi}. Specifically, if we define the equations of motion by the variation with respect to $\Phi^{\ddagger}$, we require
\begin{equation}
\label{eq:deltaL}
\delta \mathcal{L}\ =\ \bigg(\frac{\partial \mathcal{L}}{\partial \Phi}\:-\:\partial_{\alpha}\,\frac{\partial \mathcal{L}}{\partial \partial_{\alpha}\Phi}\bigg)\delta \Phi
\end{equation}
in order for the current to be conserved~\cite{Alexandre:2017foi} (for a summary, see also reference~\cite{Alexandre:2017erl}).

As an example, we can consider the global $U(1)$ transformations of the model in equation~\eqref{eq:Lag1}. The Lagrangian is invariant under the transformation $\Phi\to e^{-i\theta}\Phi$. The corresponding current
\begin{equation}
j_+^{\alpha}\ =\ i\big(\phi_1^*\partial^{\alpha}\phi_1-[\partial^{\alpha}\phi_1^*]\phi_1\big)+i\big(\phi_2^*\partial^{\alpha}\phi_2-[\partial^{\alpha}\phi_2^*]\phi_2\big)\;,
\end{equation}
however, is not conserved:
\begin{equation}
\partial_{\alpha}j^{\alpha}_+\ =\ 2i\mu^2(\phi_2^*\phi_1-\phi_1^*\phi_2)\;.
\end{equation}
On the other hand, the transformation $\Phi\to e^{-iP\theta}\Phi$, which does not leave the non-Hermitian terms in the Lagrangian invariant, leads to a conserved current
\begin{equation}
j_-^{\alpha}\ =\ i\big(\phi_1^*\partial^{\alpha}\phi_1-[\partial^{\alpha}\phi_1^*]\phi_1\big)-i\big(\phi_2^*\partial^{\alpha}\phi_2-[\partial^{\alpha}\phi_2^*]\phi_2\big)\;.
\end{equation}
Under this transformation the Lagrangian transforms to
\begin{equation}
\label{eq:Lag2}
\mathcal{L}_{\theta}\ =\ \partial_{\alpha}\phi_1^*\partial^{\alpha}\phi_1\:+\:\partial_{\alpha}\phi_2^*\partial^{\alpha}\phi_2\:-\:m_1^2|\phi_1|^2\:-\:m_2^2|\phi_2|^2\:-\:\mu^2(e^{+2i\theta}\phi_1^*\phi_2-e^{-2i\theta}\phi_2^*\phi_1)\;,
\end{equation}
and the variation $\delta \mathcal{L} = 2\mu^2(\phi_2^*\delta \phi_1-\phi_1^*\delta \phi_2)$ is consistent with equation~\eqref{eq:deltaL}. Notice, however, that the eigenspectrum \emph{is} invariant under the transformation, and the Lagrangian~\eqref{eq:Lag2} describes a one-parameter family of equivalent theories.

One can also consider the following non-Hermitian extension of the Dirac Lagrangian~\cite{Bender:2005hf}:
\begin{equation}
\mathcal{L}_{\psi}\ =\ \bar{\psi}\big(i\gamma^{\alpha}\partial_{\alpha}-m-\mu \gamma^5\big)\psi\;.
\end{equation}
The parity odd, anti-Hermitian mass term treats left- and right-handed chiralities unequally (cf.~reference~\cite{Chernodub:2017lmx}), allowing, e.g., for novel scenarios of flavour oscillations~\cite{JonesSmith:2009wy, Ohlsson:2015xsa} and neutrino mass generation~\cite{Alexandre:2015kra, Alexandre:2017fpq}, or the chiral magnetic effect to occur in equilibrium~\cite{Chernodub:2019ggz}. The conserved current is~\cite{Alexandre:2015oha}
\begin{equation}
j_{\psi}^{\alpha}\ =\ \bar{\psi}\gamma^{\alpha}\bigg(1+\frac{\mu}{m}\,\gamma^5\bigg)\psi\;,
\end{equation}
corresponding to the transformation $\psi\to e^{-i\theta(1+\mu\gamma^5/m)}\psi$, for which the variation $
\delta \mathcal{L}_{\psi}=-\,2\mu\bar{\psi}\gamma^5\delta \psi$ is consistent with choosing the equations of motion by varying with respect to $\bar{\psi}$~\cite{Alexandre:2017foi}.

\section{Global symmetries and the Goldstone theorem}
\label{sec:Goldstone}

The proof of the Goldstone theorem~\cite{Nambu:1960xd, Goldstone:1961eq, Goldstone:1962es} relies on the existence of a conserved current (see, e.g., reference~\cite{Weinberg}), and it should therefore hold also in the non-Hermitian case.

We consider the following theory with a spontaneously broken global $U(1)$ symmetry~\cite{Alexandre:2018uol}:
\begin{equation}
\label{eq:Lag1}
\mathcal{L}\ =\ \partial_{\alpha}\phi_1^*\partial^{\alpha}\phi_1\:+\:\partial_{\alpha}\phi_2^*\partial^{\alpha}\phi_2\:+\:m_1^2|\phi_1|^2\:-\:m_2^2|\phi_2|^2\:-\:\mu^2(\phi_1^*\phi_2-\phi_2^*\phi_1)\:-\:\frac{g}{4}\,|\phi_1|^4\;.
\end{equation}
Choosing the equations of motion to be defined by the variation with respect to $\Phi^{\ddagger}$, as before, the symmetry-breaking vacua of the theory are given by the solutions to~\cite{Alexandre:2018uol}
\begin{equation}
\bigg[\frac{g}{2}\,|\phi_1|^2\phi_1\:-\:m_1^2\phi_1\:+\:\mu^2\phi_2\bigg]_{\phi_{1,2}\,=\,v_{1,2}}\ =\ 0\qquad \text{and} \qquad
\bigg[m_2^2\phi_2\:-\:\mu^2\phi_1\bigg]_{\phi_{1,2}\,=\,v_{1,2}}\ = \ 0\;.
\end{equation}
Up to an overall complex phase, we find
\begin{equation}
\label{eq:vevs}
\begin{pmatrix}
v_1\\ v_2
\end{pmatrix}
\ =\ \sqrt{2\,\frac{m_1^2m_2^2-\mu^4}{gm_2^2}}\begin{pmatrix} 1\\ \frac{\mu^2}{m_2^2}\end{pmatrix}\;.
\end{equation}
Notice that $v_2$ depends on the sign of $\mu^2$, and the spontaneous breaking of the $\mathcal{PT}$ symmetry manifests in non-$\mathcal{PT}$-symmetric terms in the action at linear order in the fluctuations~\cite{Alexandre:2018uol}.

The equations of motion for the fluctuations $\hat{\phi}_{1,2}=\phi_{1,2}-v_{1,2}$ read~\cite{Alexandre:2018uol}
\begin{equation}
\Box\begin{pmatrix} \hat{\phi}_1 \\ \hat{\phi}_1^*\\ \hat{\phi}_2 \\ \hat{\phi}_2^*\end{pmatrix}\ =\ \begin{pmatrix} \frac{m_1^2m_2^2-2\mu^4}{m_2^2} & \frac{m_1^2m_2^2-\mu^4}{m_2^2} & \mu^2 & 0\\ \frac{m_1^2m_2^2-\mu^4}{m_2^2} & \frac{m_1^2m_2^2-2\mu^4}{m_2^2} & 0 & \mu^2 \\ -\,\mu^2 & 0 & m_2^2 & 0 \\ 0 &-\,\mu^2 & 0 & m_2^2\end{pmatrix}\begin{pmatrix} \hat{\phi}_1 \\ \hat{\phi}_1^*\\ \hat{\phi}_2 \\ \hat{\phi}_2^*\end{pmatrix}+\mathcal{O}(\hat{\phi}^2)\;.
\end{equation}
The eigenspectrum depends only on $\mu^4$, as before, and it remains real for $(2m_1^2m_2^2-3\mu^4-m_2^4)^2\geq4\mu^4m_2^4$ and positive semi-definite for $\mu^4<m_2^4$ (when $m_1^2>m_2^2$). Moreover, it contains one zero eigenvalue, corresponding to the Goldstone mode
\begin{equation}
G\ \propto\ {\rm Im}\,\hat{\phi}_1\:-\:\frac{\mu^2}{m_2^2}\,{\rm Im}\,\hat{\phi}_2\;,
\end{equation}
consistent with the Goldstone theorem, as we could have confirmed directly from the conserved current (see reference~\cite{Alexandre:2018uol}). As pointed out in reference~\cite{Mannheim:2018dur}, the Goldstone mode is normalisable with respect to the $\mathcal{PT}$ inner product only away from the exceptional point, which lies at $\mu^2=\pm\,m_2^2$, when the mass matrix above becomes defective and we lose an eigenvector.

\section{Local symmetries and the Englert-Brout-Higgs mechanism}

We turn our attention now to the case of spontaneously broken local symmetries in non-Hermitian theories (see reference~\cite{Alexandre:2018xyy}). Motivated by the fact that \smash{$\partial_{\alpha}\partial_{\beta}F^{\alpha\beta}$} vanishes identically due to the antisymmetry of the field-strength tensor \smash{$F^{\alpha\beta}=\partial^{\alpha}A^{\beta}-\partial^{\beta}A^{\alpha}$}, we might be tempted to gauge the $U(1)$ symmetry of the model of sections~\ref{sec:model} and \ref{sec:Noether} by minimally coupling a gauge field $A_{\alpha}$ to the \emph{conserved} current via the covariant derivative \smash{$\mathcal{D}_{\alpha}=\mathbb{I}_2\partial_{\alpha}+iqPA_{\alpha}$} of the complex doublet $\Phi$. Proceeding in this way, we obtain the Lagrangian
\begin{equation}
\mathcal{L}_-\ =\ [D^+_{\alpha}\phi_1]^*D^{\alpha}_+\phi_1\:+\:[D^-_{\alpha}\phi_2]^*D_-^{\alpha}\phi_2\:-\:m_1^2|\phi_1|^2\:-\:m_2^2|\phi_2|^2\:-\:\mu^2(\phi_1^*\phi_2-\phi_2^*\phi_1)\:-\:\frac{1}{4}\,F_{\alpha\beta}F^{\alpha\beta}\;,
\end{equation}
where $D^{\alpha}_{\pm}=\partial^{\alpha}\pm iq A^{\alpha}$. The conserved current is
\begin{equation}
j_{A,-}^{\alpha}\ =\ iq\big(\phi_1^*D^{\alpha}_+\phi_1-[D^{\alpha}_+\phi_1]^*\phi_1\big)\:-\:iq\big(\phi_2^*D^{\alpha}_-\phi_2-[D^{\alpha}_-\phi_2]^*\phi_2\big)\;,
\end{equation}
corresponding to the transformations
\begin{equation}
\Phi(x)\ \rightarrow\ e^{-iqPf(x)}\Phi(x)\qquad \text{and}\qquad A^{\alpha}(x)\ \rightarrow\ A^{\alpha}(x)\:+\:\partial^{\alpha}f(x)\;.
\end{equation}
However, these transformations do not leave the non-Hermitian part of the Lagrangian invariant. This loss of gauge invariance, while not affecting the classical scalar eigenspectrum, leads to a (problematic) non-transverse one-loop polarisation tensor~\cite{Alexandre:2018xyy}:
\begin{equation}
k_{\alpha}\Pi^{\alpha\beta}(k^2=0)\ =\  \frac{q^2}{8\pi^2}\,\frac{k^{\beta}\mu^4}{\big(M_+^2-M_-^2\big)^3}\bigg[M_+^4-M_-^4+2M_+^2M_-^2\ln\bigg(\frac{M_-^2}{M_+^2}\bigg)\bigg]\;.
\end{equation}

We can restore gauge invariance in the weak sense by going beyond the minimal coupling prescription and modifying the non-Hermitian terms to include factors of the Wilson line~\cite{Wilson:1974sk}
\begin{equation}
W(x)\ \equiv\ \exp\bigg[iq\int^x\!{\rm d}y^{\alpha}\;A_{\alpha}\bigg]\;,
\end{equation}
where the path starts at the boundary (infinity) and ends at $x$. The Lagrangian takes the form~\cite{Alexandre:2018xyy}
\begin{align}
\label{eq:wilsonlag}
\mathcal{L}_W\ &=\ [D^+_{\alpha}\phi_1]^*D^{\alpha}_+\phi_1\:+\:[D^-_{\alpha}\phi_2]^*D^{\alpha}_-\phi_2\:-\:m_1^2|\phi_1|^2\:-\:m_2^2|\phi_2|^2\:-\:\mu^2(W^{*2}\phi_1^*\phi_2-W^{2}\phi_2^*\phi_1)\nonumber\\&\qquad-\:\frac{1}{4}\,F_{\alpha\beta}F^{\alpha\beta}\;.
\end{align}
The Wilson line transforms as \smash{$W(x)\to W(x)e^{iqf(x)}$} for gauge transformations that vanish at infinity, and the Lagrangian in equation~\eqref{eq:wilsonlag} is gauge invariant. However, we have restored gauge invariance at the cost of introducing a path dependence~\cite{Alexandre:2018xyy}. Moreover, the direct coupling to the non-Hermitian term may, in general, be inconsistent with the reality of the gauge field.

Alternatively, we can couple minimally to the \emph{non-conserved} current, assigning like charges to the complex scalar fields $\phi_1$ and $\phi_2$. In this case, however, the Maxwell equation is inconsistent unless we can extend the gauge Lagrangian in an appropriate way. It turns out that a sufficient ingredient is the usual gauge-fixing term~\cite{Alexandre:2018xyy}, and we take
\begin{align}
\mathcal{L}_+\ &=\ [D_{\alpha}\phi_1]^*D^{\alpha}\phi_1\:+\:[D_{\alpha}\phi_2]^*D^{\alpha}\phi_2\:-\:m_1^2|\phi_1|^2\:-\:m_2^2|\phi_2|^2\:-\:\mu^2(\phi_1^*\phi_2-\phi_2^*\phi_1)\nonumber\\&\qquad-\:\frac{1}{4}\,F_{\alpha\beta}F^{\alpha\beta}\:-\:\frac{1}{2\xi}\,(\partial_{\alpha}A^{\alpha})^2\;,
\end{align}
with $D^{\alpha}\equiv D^{\alpha}_+$ and the gauge symmetry restricted to transformations involving gauge functions that satisfy $\Box f =0$. The Maxwell equation becomes
\begin{equation}
\Box A^{\alpha}\:-\:(1-1/\xi)\partial^{\alpha}\partial_{\beta}A^{\beta}\ =\ iq\big(\phi_1^*D^{\alpha}\phi_1-[D^{\alpha}\phi_1]^*\phi_1\big)\:+\:iq\big(\phi_2^*D^{\alpha}\phi_2-[D^{\alpha}\phi_2]^*\phi_2\big)\ = \ j_{A,+}^{\alpha}\;,
\end{equation}
and its divergence leads to the constraint
\begin{equation}
\label{eq:boxpi}
\Box \pi_0\ = \ 2iq\mu^2(\phi_1^*\phi_2-\phi_2^*\phi_1)\;,
\end{equation}
where $\pi_0=-\,\partial_{\alpha}A^{\alpha}/\xi$ is the conjugate momentum to $A_0$. This precludes the usual Lorenz gauge condition $\partial_{\alpha} A^{\alpha}=0$. (While $\phi_1^*\phi_2-\phi_2^*\phi_1$ may vanish classically, we have $\braket{\phi_1^*\phi_2}-\braket{\phi_2^*\phi_1}\neq 0$.)

Having defined a consistent non-Hermitian deformation of scalar quantum electrodynamics, we are now in a position to consider the extension of the Englert-Brout-Higgs mechanism~\cite{Englert:1964et, Higgs:1964ia, Higgs:1964pj} to non-Hermitian Abelian theories. Taking the Lagrangian
\begin{align}
\mathcal{L}\ &=\ [D_{\alpha}\phi_1]^*D^{\alpha}\phi_1\:+\:[D_{\alpha}\phi_2]^*D^{\alpha}\phi_2\:+\:m_1^2|\phi_1|^2\:-\:m_2^2|\phi_2|^2\:-\:\mu^2(\phi_1^*\phi_2-\phi_2^*\phi_1)\:-\:\frac{g}{4}\,|\phi_1|^4\nonumber\\&\qquad-\:\frac{1}{4}\,F_{\alpha\beta}F^{\alpha\beta}\:-\:\frac{1}{2\xi}\,(\partial_{\alpha}A^{\alpha})^2\;,
\end{align}
we have the same symmetry-breaking vacuum as the global case in section~\ref{sec:Goldstone}. Expanding around this vacuum, equation~\eqref{eq:vevs}, we find that the gauge field obtains a mass
\begin{equation}
M_A^2\ =\ 2q^2(|v_1|^2+|v_2|^2)\;,
\end{equation}
such that we can indeed generate a gauge-invariant vector boson mass via a non-Hermitian extension of the Englert-Brout-Higgs mechanism~\cite{Alexandre:2018xyy} (cf.~reference~\cite{Mannheim:2018dur}). The generalisation to the non-Abelian case~\cite{Kibble:1967sv} may be presented elsewhere.

\section{Concluding remarks}

We have discussed global and local symmetries in the context of a non-Hermitian complex-scalar model that exhibits $\mathcal{PT}$ symmetry. We have shown that a careful treatment of Noether's theorem indicates that there exist conserved currents for non-Hermitian models but that these correspond to transformations that do not leave the Lagrangian invariant. In the case of gauge symmetries, we have argued that it is necessary to couple to the non-conserved current in order to preserve gauge invariance but that the non-Hermitian nature of the theory leads to a constraint on the gauge field, precluding the Lorenz gauge condition. In the case of spontaneously broken global and local symmetries, we have illustrated that the Goldstone theorem and the Englert-Brout-Higgs mechanism are still borne out. These results pave the way for further studies aiming to construct consistent non-Hermitian extensions of the Standard Model of particle physics.

\ack

The work of PM is supported by a Leverhulme Trust Research Leadership Award. PM would like to thank Jean Alexandre, Carl Bender, John Ellis and Dries Seynaeve for their collaboration in this area, and Maxim Chernodub, Alberto Cortijo, Philip Mannheim and the participants of the $\mathcal{PT}$ symmetry track of DISCRETE 2018 for enjoyable and enlightening discussions.

\end{document}